\documentclass[aps,prl,reprint,superscriptaddress]{revtex4-2}

\usepackage{amsmath} 
\usepackage{amssymb}
\usepackage{mathrsfs}
\usepackage{graphicx} 
\usepackage{float}
\usepackage{helvet}
\usepackage{makecell}
\usepackage{multirow}
\usepackage{siunitx}
\usepackage{booktabs} 
\usepackage{makecell} 
\usepackage[colorlinks=true,linkcolor=blue,filecolor=magenta,
urlcolor=blue]{hyperref}

\usepackage{xr}       

\renewcommand{\arraystretch}{1.25}

\usepackage{soul}
\usepackage[dvipsnames]{xcolor}

\usepackage[version=4]{mhchem}

\begin{document}

\title{Real-Space Mapping of Electronic Conductivity in Complex Materials}

\author{C. Ugwumadu}
\email{cugwumadu@lanl.gov}
\affiliation{Quantum \& Condensed Matter (T-4) Group, Los Alamos National Laboratory, Los Alamos, NM, USA}

\author{D. A. Drabold}
\affiliation{Department of Physics and Astronomy, Ohio University, Athens, OH, USA}

\author{R. M. Tutchton}
\affiliation{Quantum \& Condensed Matter (T-4) Group, Los Alamos National Laboratory, Los Alamos, NM, USA}

\begin{abstract}
We introduce KuboMap, a real-space representation of electronic conductivity derived from the Kubo--Greenwood formula. KuboMap defines a nonnegative conductivity density whose spatial integral recovers the total conductivity and whose form is guided by Mott's picture of transport through spatially overlapping electronic states. This construction provides a direct map of the transport-active regions of a material. Applied to aluminum, KuboMap recovers an extended metallic conduction network. Applied to amorphous silicon, it distinguishes an insulating defect-free network from a defective structure in which localized near-Fermi states form connected hopping-like pathways. In silicon-oxides, it captures the loss of conduction as increasing oxygen content disrupts Silicon-rich transport networks. KuboMap provides a physically transparent route from Kubo--Greenwood conductivity to real-space transport pathways in complex materials.
\end{abstract}


\maketitle

Electronic conductivity is a fundamental observable in condensed-matter physics and materials science, yet in complex materials, its microscopic origin is often difficult to interpret. In practical electronic-structure calculations, conductivity is commonly evaluated from the Kubo--Greenwood formula \cite{Kubo1957,Greenwood1958}, which yields a global response but does not directly reveal how conduction occurs in real space. This limitation is especially relevant in disordered and heterogeneous systems, where transport depends strongly on the spatial organization and localization of states near the Fermi level.

For non-crystalline materials, Anderson and Mott emphasized that transport is not simply controlled by extended Bloch states, but by localized states near the band edges and by the possibility of hopping between states that are both energetically accessible and spatially overlapping \cite{Anderson1958,Mott1969,Mott1968}. This suggests that a useful real-space representation of conductivity should be built from pairs of states rather than from isolated orbitals. Existing approaches, such as the space-projected conductivity (SPC) method \cite{Subedi2020} and Hindley--Mott-inspired \(N^2\) constructions \cite{Nepal2025_N2,HusseinTAHM2026}, made progress in connecting conductivity to local structure. 

Here we introduce \emph{KuboMap}, which we regard as a conceptual shift in how conductivity is represented in nanostructures. The central idea is to decompose conductivity into normalized pair densities constructed from products of Kohn--Sham (KS) eigenstates. The resulting spatial field reproduces the total conductivity upon spatial integration while identifying the real-space regions where transport-active state pairs coexist. We show that KuboMap recovers extended metallic conduction in aluminum, distinguishes insulating and defect-mediated transport regimes in amorphous silicon, and captures progressive suppression of conduction with increasing oxygen content in silicon-oxides.

In the energy eigenbasis, \(\{|m\rangle\}\), satisfying \(\hat H|m\rangle=E_m|m\rangle\), where \(\hat H\) denotes the KS Hamiltonian, we start from a Kubo--Greenwood expression for the diagonal conductivity along the Cartesian direction \(\alpha\in\{x,y,z\}\) \cite{Bose1993},
\begin{equation}
\sigma_{\alpha \alpha}^{(\eta)}
=
\sum_{mn}\gamma_{mn}^{\alpha \alpha}w_m w_n,
\label{eq:sigma}
\end{equation}
with
\begin{equation}
\gamma_{mn}^{\alpha \alpha}
=
\frac{2\pi e^2\hbar}{\Omega}\,
\left|\left\langle n \left| \frac{\hat P^\alpha}{\mu} \right| m  \right\rangle\right|^2,
\label{eq:gamma}
\end{equation}
and Gaussian broadening weights
\begin{equation}
w_m
=
\frac{1}{\sqrt{2\pi}\eta}
\exp\!\left[-\frac{(E_m-E_F)^2}{2\eta^2}\right].
\label{eq:w}
\end{equation}
Here \(\Omega\) is the cell volume, \(\mu\) is the electron mass, $\hbar$ is Planck's constant, $E_F$ is the Fermi energy, and \(\hat P^\alpha\) is the \(\alpha\)-th Cartesian component of the momentum operator. For a finite supercell, \(\eta\) accounts for both finite-size level spacing and thermal broadening; details of its choice are discussed later.

\begin{table*}[!t]
\scriptsize
\setlength{\tabcolsep}{12pt}
\renewcommand{\arraystretch}{1.4}
\centering
\caption{Structural broadening (\(\eta_s\)), thermal broadening (\(\eta_t\)), resulting Gaussian broadening parameter (\(\eta\)), and conductivity \(\sigma^{(\eta)}\) for all structures at 300 K. Broadening parameters are reported in eV, and conductivities are reported in S/m.}
\label{tab:eta_values}
\begin{tabular}{l|ccccccc}
\hline
 & $c$-Al & $a$-Si-216 & $a$-Si-512 & SiO$_{1.3}$ & SiO$_{1.5}$& SiO$_{1.7}$ & SiO$_{2}$\\
\hline
\(\eta_s\)  & 0.026 & 0.04 & 0.03 & 0.09 & 0.106 & 0.118 & 0.135 \\
\(\eta_t\)  & 0.008 & 0.042 & 0.02 & 0.097 & 0.146 & 0.194 & 0.196 \\
\(\eta\)  & 0.035 & 0.085 & 0.05 & 0.20 & 0.255 & 0.32 & 0.335 \\
\(\sigma^{(\eta)}\)  & $3.5 \times 10^5$ & $2.6 \times 10^{-5}$ &  $498$ & $282$ & $0.26$ & $0.02$ & $4.6\times 10^{-8}$ \\
\hline
\end{tabular}
\end{table*}

Motivated by Mott's \cite{Mott1969} and Hindley's \cite{Hindley1970_RP} argument that $\sigma \propto N^2(E_F)$ \cite{Nepal2025_N2}, where $N(E_F)$ is the density of states at the Fermi level, we seek a nonnegative spatial density $\rho_{\alpha \alpha}^{(\eta)}(\mathbf r)$ whose integral recovers Eq.~\eqref{eq:sigma}. A natural choice is to build each $(m,n)$ contribution from the product of the local probability densities $|\psi_m(\mathbf r)|^2$ and $|\psi_n(\mathbf r)|^2$, for KS states $\langle \mathbf{r}|m\rangle$ and $\langle \mathbf{r}|n\rangle$ at position $\mathbf r$. First, we define the overlap matrix 
\begin{equation}\label{eq:beta}
\beta_{mn}^{-1} = \int d^3r\;|\psi_m(\mathbf r)|^2\,|\psi_n(\mathbf r)|^2,
\end{equation}
and construct the KuboMap conductivity density
\begin{align}\label{eq:rho}
\rho_{\alpha \alpha}^{(\eta)}(\mathbf r)
&=
\sum_{m,n}
\gamma_{mn}^{\alpha \alpha}\,\beta_{mn}\,
|\psi_m(\mathbf r)|^2 |\psi_n(\mathbf r)|^2\;w_m w_n, \\
\rho^{(\eta)}(\mathbf r)&=
\frac{1}{3}\sum_{\alpha\in\{xyz\}}\rho_{\alpha \alpha}^{(\eta)}(\mathbf r).
\end{align}
By construction:
\begin{equation}\label{eq:betanomalization}
\int d^3r\;\beta_{mn}\,|\psi_m(\mathbf r)|^2\,|\psi_n(\mathbf r)|^2 = 1,
\end{equation}
from which
\begin{align}
    \sigma_{\alpha \alpha}^{(\eta)} =& \int d^3r\,\rho_{\alpha \alpha}^{(\eta)}(\mathbf r) \\ \sigma^{(\eta)}=& \frac{1}{3}\sum_{\alpha\in\{xyz\}}\sigma_{\alpha \alpha}^{(\eta)}
\label{eq:sumrule}
\end{align}
Equation~\eqref{eq:rho} therefore defines a nonnegative spatial conductivity density whose integral recovers the Kubo--Greenwood conductivity [Eq. \eqref{eq:sigma}]. A formal interpretation of Eq. \eqref{eq:rho} is provided in Sec.~\textcolor{magenta}{S1} of the supplementary material (SM) \cite{SM}, where $|\psi_m(\mathbf r)|^2|\psi_n(\mathbf r)|^2$ is represented as the diagonal kernel of a positive operator on the doubled Hilbert space $\mathcal H\otimes\mathcal H$. 

\begin{figure}[h]
    \centering
    \includegraphics[width=\linewidth]{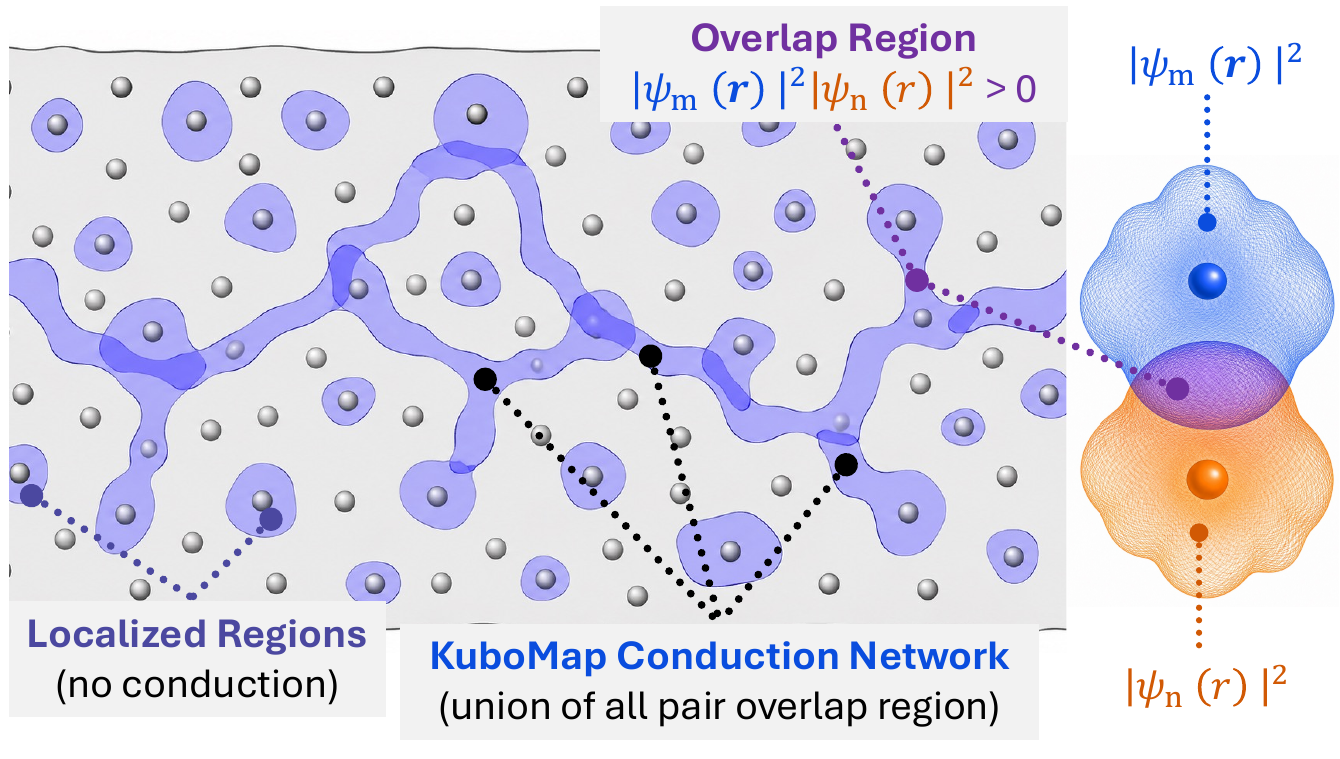}
    \caption{Geometric interpretation of KuboMap.}
    \label{fig:Figure1}
\end{figure}

The physical interpretation of Eq.~\eqref{eq:rho} is illustrated schematically in Fig.~\ref{fig:Figure1}. Individual near-\(E_F\) states may be spatially localized and therefore do not, by themselves, define a conduction pathway (isolated purple blobs). By contrast, the products \( |\psi_m(\mathbf r)|^2 |\psi_n(\mathbf r)|^2 \) isolate the regions where two relevant states coexist in space, and the union of these pair-overlap regions defines the conducted transport network identified by KuboMap (connected purple channels). A mathematical description of this picture is given in Sec. \textcolor{magenta}{S2} of SM \cite{SM}.

 \begin{figure*}[!t]
    \centering
    \includegraphics[width=\textwidth]{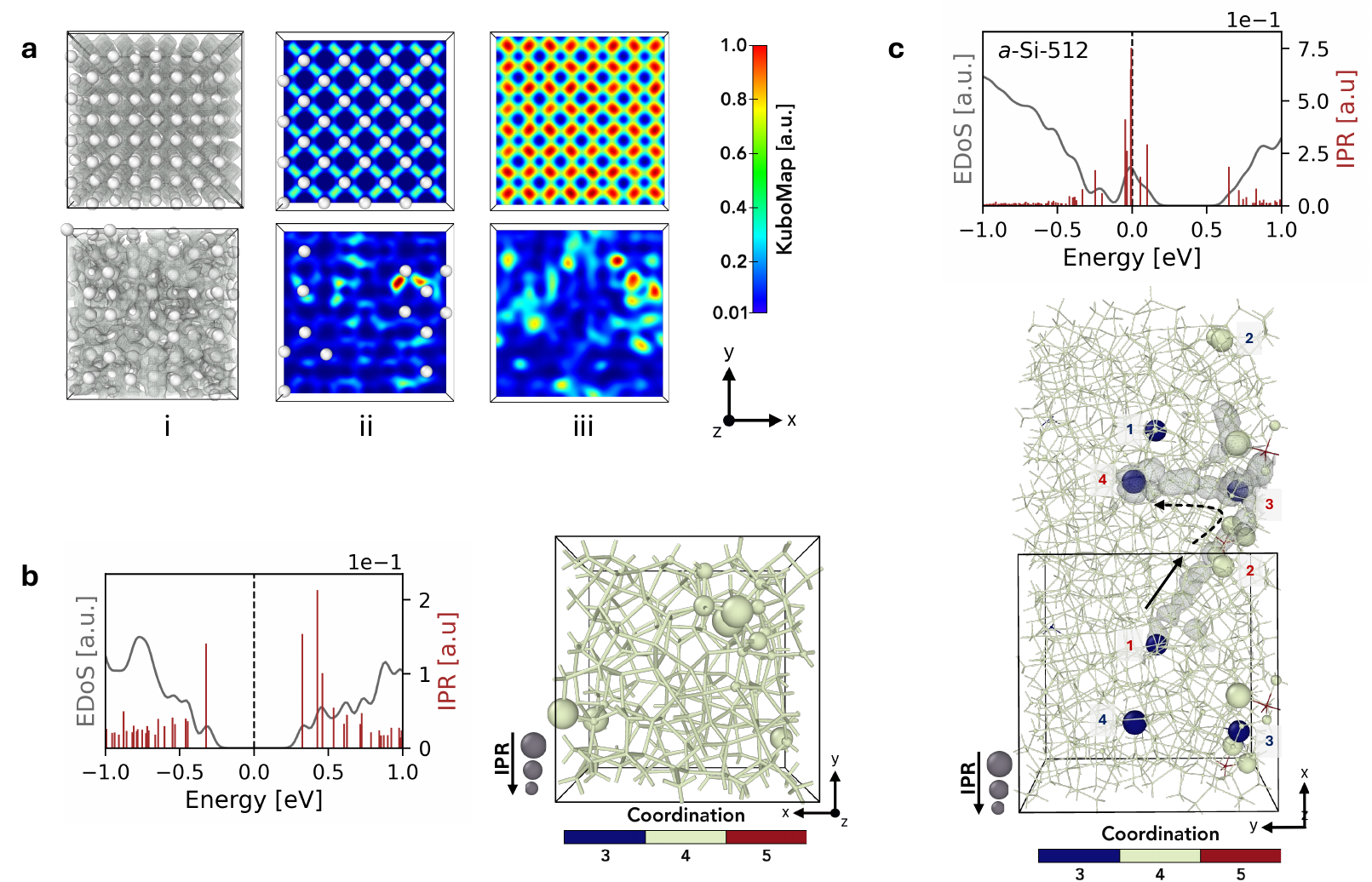}
    \caption{KuboMap in aluminum (Al) and amorphous silicon ($a$-Si). (a) KuboMap for crystalline Al (top row) and thermally disordered Al (bottom row). Column (i) shows the bulk KuboMap distribution, column (ii) an \(xy\) slice through an atomic lattice plane at \(z=12~\text{\AA}\), and column (iii) an \(xy\) slice through an interstitial region at \(z=13~\text{\AA}\). (b) Electronic density of states (EDoS; gray) and inverse participation ratio (IPR; red) for \(a\)-Si-216, together with its atomic structure, which exhibits no connected KuboMap pathway. (c) EDoS (gray) and IPR (red) for \(a\)-Si-512, and the corresponding KuboMap projection, shown as a gray isosurface on the atomic structure. Repetition of the supercell along [100] reveals a connected transport pathway and selected atoms in the original and repeated cells are labeled 1--4: red labels indicate atoms along the conduction pathway, while blue labels mark corresponding atoms across the periodic image.}
    \label{fig:Figure2}
\end{figure*}

We applied KuboMap to face-centered cubic (FCC) aluminum (\(c\)-Al), amorphous silicon (\(a\)-Si), and silicon-oxide stoichiometries (SiO$_x$). The calculations were performed using density functional theory within the Vienna \emph{ab initio} Simulation Package (VASP) \cite{VASP}. We used projector-augmented-wave potentials \cite{PAW} and the Perdew--Burke--Ernzerhof exchange-correlation functional \cite{PBE}.

For $c$-Al, we used a 256-atom model obtained from Ref.~\cite{HusseinTAHM2026}. For $a$-Si, we considered two models: a 216-atom model ($a$-Si-216) from Ref.~\cite{DTW}, generated using the Wooten--Winer--Weaire approach and exhibiting a clean electronic gap, and a 512-atom model ($a$-Si-512) from Ref.~\cite{Morrow2024}, obtained from machine-learning-driven molecular-dynamics (MD) simulations and containing coordination defects and near-\(E_F\) states. The silicon-oxide models, SiO$_x$ with stoichiometries \(x = 1.3, 1.5, 1.7,\) and \(2.0\), were taken from Ref.~\cite{Ugwumadu2023_Siox}.

For each structure, we averaged over 10 snapshots sampled at 300 K from a 5 ps trajectory with a 0.5 fs timestep, following 1 ps of equilibration, to estimate the room-temperature conductivity. Temperature-dependent conductivity for these models will be addressed in future work, together with calculations using hybrid functionals, such as the Heyd--Scuseria--Ernzerhof functional \cite{HSE2023}, to assess the eigenvalue placement and localization.

KuboMap was evaluated on a uniform grid with 0.2 \AA~spacing for all systems, using 30 states above and 30 states below the Fermi level. Additional implementation details are provided in Sec.~\textcolor{magenta}{S3} of the SM \cite{SM}.

The Gaussian broadening parameter \(\eta\) [Eq. \eqref{eq:w}] must account for both finite-size spectral discreteness in the KS spectrum and thermal fluctuations of the near-\(E_F\) states. We therefore require
\begin{equation}\label{eq:eta_components}
\eta^2 \ge \eta_s^2+\eta_t^2,
\end{equation}
where \(\eta_s\) is the structural contribution, estimated from the spacing of the states nearest \(E_F\), and \(\eta_t\) is the thermal contribution, estimated from the root-mean-square (rms) fluctuation of the same states along the MD trajectory \cite{AllenFeldman1993,Drabold1991}. We discuss this further in Sec. \textcolor{magenta}{S4} \cite{SM}.

Table~\ref{tab:eta_values} summarizes the average values of $\eta_s$, $\eta_t$, $\eta$, and the corresponding average conductivity $\sigma^{(\eta)}$ evaluated at $\eta$ over the 10 snapshots for each structure. The individual snapshot values are discussed in Sec. \textcolor{magenta}{S5} of the SM \cite{SM}. The estimates of $\eta_s$ and $\eta_t$ used 30 states below and 30 states above the Fermi level. The corresponding near-$E_F$ states and their rms fluctuations, used to determine $\eta_s$ and $\eta_t$, are shown in Fig. \textcolor{magenta}{S1} of the SM \cite{SM}.

In Table~\ref{tab:eta_values}, clear material-dependent trends emerge for the Gaussian broadening parameters. Thermally disordered $c$-Al has the smallest broadening parameters due to its dense metallic spectrum near \(E_F\) [Fig. \textcolor{magenta}{S1}(a/d)]. The amorphous-Si models show larger broadening scales. The values for \(\eta_s\) and \(\eta_t\) are comparable in \(a\)-Si-216, which reflects the strong sensitivity of the sparse near-gap spectrum to thermal fluctuations [Fig. \textcolor{magenta}{S1}(b/d)]. The $a$-Si-512 model contains additional states near $E_F$, associated with non-four-fold coordinated Si atoms as discussed in Ref.~\cite{Morrow2024}; consequently, its values are smaller than that of $a$-Si-216 [Fig. \textcolor{magenta}{S1}(c/d)].

The SiO$_x$ series exhibits the largest broadening parameters overall, with both \(\eta_s\) and \(\eta_t\) increasing systematically as the oxygen content rises from SiO$_{1.3}$ to SiO$_2$. This trend reflects the progressive depletion of near-\(E_F\) states and the growing sensitivity of the electronic-structure to thermal fluctuations as the system approaches stoichiometric SiO$_2$ [Fig. \textcolor{magenta}{S1}(e--i)]. Accordingly, the selected \(\eta\) values are therefore largest for the oxide models, ensuring that the broadening exceeds both structural and thermal lower bounds while remaining physically guided.

Figure~\ref{fig:Figure2}(a) shows KuboMap for crystalline and thermally disordered FCC Al. In both cases the conductivity density is spatially extended, consistent with metallic transport [Fig.~\ref{fig:Figure2}(a,i)]. The crystalline system exhibits a regular, spatially periodic KuboMap pattern, whereas the thermally disordered structure shows a more heterogeneous distribution, reflecting modulation of the local overlap by atomic displacements.

The contrast is more apparent in the \(xy\) slices. In the lattice-plane slice at \(z=12~\text{\AA}\) [Fig.~\ref{fig:Figure2}(a,ii)], crystalline Al displays a regular pattern tied to the ordered atomic arrangement, while thermal disorder partially disrupts this periodicity and redistributes the local intensity. In the interstitial region slice at \(z=13~\text{\AA}\) [Fig.~\ref{fig:Figure2}(a,iii)], crystalline Al forms a highly regular interstitial network, whereas the thermally disordered case is more diffuse and irregular. In both cases, however, the KuboMap remains connected across the cell, consistent with metallic transport.

The conductivity of the thermally disordered $c$-Al at $\eta =$ 0.05 eV is $3.5 \times 10^5$ S/m, typical of a conducting metal (Table \ref{tab:eta_values}) \cite{Subedi2022_Al}. The conductivity of crystalline Al is not reported, since the DC conductivity of FCC Al is infinite. Additional details for $c$-Al are given in Sec. \textcolor{magenta}{S6} of the SM \cite{SM}. %

Figures~\ref{fig:Figure2}(b) and (c) compare two limiting cases for amorphous silicon: defect-free \(a\)-Si-216, which exhibits a clean gap around \(E_F\), and defective \(a\)-Si-512, which contains states near \(E_F\). In both structures, localized states appear in the valence- and conduction-band tails, and their degree of localization is quantified by the inverse participation ratio (IPR), 
    \(\mathcal I_n=
    \tfrac{\sum_a c_{n;a}^2}{\left[\sum_a c_{n;a}\right]^2}\),
where \(c_{n;a}\) is the square of the wavefunction coefficient of KS state \(n\) from the atomic orbital \(a\) (s, p, and d).  \(\mathcal I_n  \rightarrow 0\) corresponds to extended states, while \(\mathcal I_n \rightarrow 1\) indicates localization. The essential difference is whether such KS states occur close enough to \(E_F\) to support electronic transport pathway.
 

In $a$-Si-216, the electronic density of states (EDoS), in the left panel of Fig.~\ref{fig:Figure2}(b), shows a clear gap around $E_F$ (gray curve) with Fermi level at 0 eV (black dashed line), consistent with insulating behavior. Although some band-edge states exhibit finite IPR (red lines), the absence of appreciable near-$E_F$ spectral weight implies that few states are available to support electronic transport. Accordingly, no KuboMap conduction pathway is observed in the structure as shown in the right panel of Fig.~\ref{fig:Figure2}(b), where the atomic radii are scaled by \(\mathcal I_n\) and color-coded by coordination number. 

In $a$-Si-512, The EDoS/IPR plot in the top panel of Fig.~\ref{fig:Figure2}(c) reveal that the Fermi level is closer to the valence-band edge (reminiscent of a $p$-type doping network), with several states near $E_F$. In particular, the two closest KS states above $E_F$ have small splittings of $\approx 0.054$ eV and $\approx 0.045$ eV, whereas the third closest state is much farther away, with $\approx 0.55$ eV (also see Fig. \textcolor{magenta}{S1}(c) \cite{SM}). These near-$E_F$ states are associated with coordination defects, including non-four-fold coordinated Si atoms (see discussion in Ref. \cite{Morrow2024}).

Unlike \(a\)-Si-216, \(a\)-Si-512 exhibits a connected KuboMap pathway, shown in the bottom panel of Fig.~\ref{fig:Figure2}(c). The gray isosurface identifies the regions that dominate the electronic conduction pathway. Repeating the supercell once along [100] shows that this pathway is not confined to isolated defect sites, but remains connected across the periodically repeated structure. The labeled atoms (1--4) mark how the channel crosses the supercell boundary and continues into the image cell. The pathway passes through atoms with large \(\mathcal I_n\) weight (scaled by atomic radii), including both under-coordinated and over-coordinated atoms (in blue and red), indicating that conduction in defective \(a\)-Si is concentrated along structural motifs that support localized near-\(E_F\) states.

The contrast in conductivity between the two amorphous Si models is correspondingly strong (Table \ref{tab:eta_values}). For insulators like \(a\)-Si-216, any physically reasonable choice of \(\eta\) gives essentially zero conductivity; at the selected value \(\eta=0.085\) eV, we obtain only \(\sigma=2.6\times10^{-5}\) S/m, which is consistent with experimentally reported values at room-temperature \cite{MIT-Mat_database,Lewis1972}. By contrast, defective \(a\)-Si-512 is conductive with  \(\sigma \approx 498\) S/m at \(\eta=0.05\) eV, implying that defect-induced near-\(E_F\) localized KS states strongly supports hopping-like transport~\cite{Anderson1958, Mott1969}. Additional details for amorphous Si are given in Sec. \textcolor{magenta}{S7} \cite{SM}.

 \begin{figure*}[!t]
    \centering
    \includegraphics[width=\textwidth]{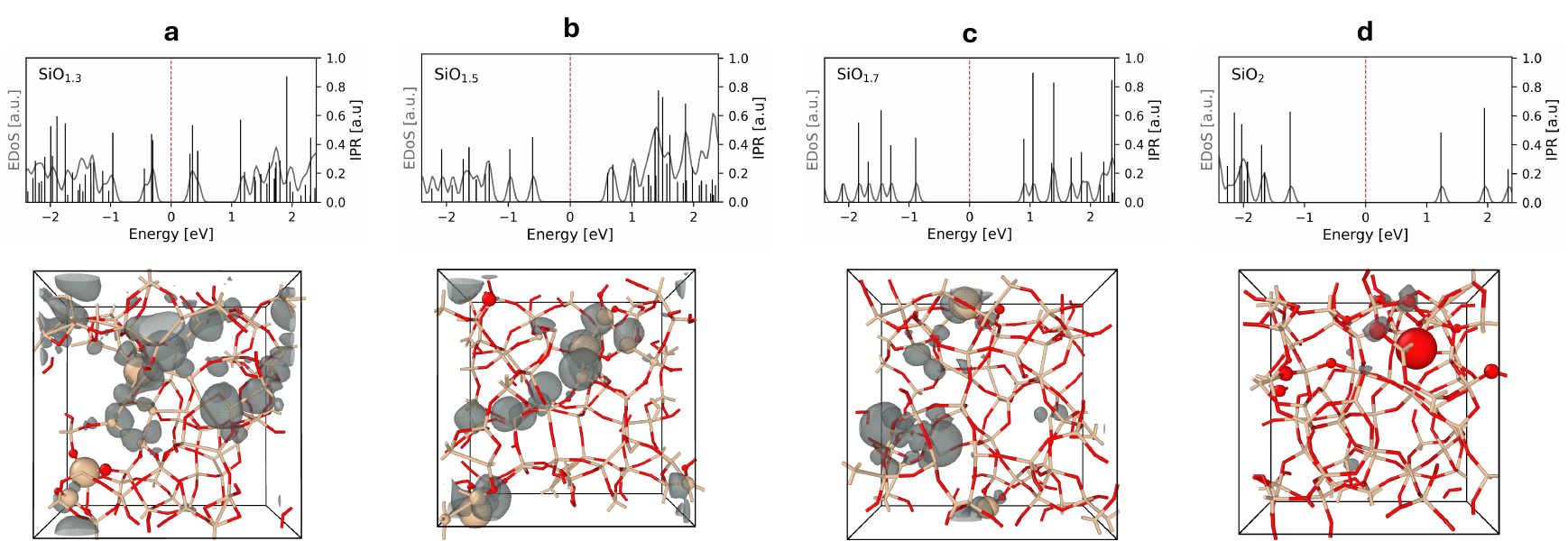}
    \caption{Analysis of silicon-oxide systems. The top panels show the electronic density of states (gray) and inverse participation ratio (IPR; black) for (a) SiO$_{1.3}$, (b) SiO$_{1.5}$, (c) SiO$_{1.7}$, and (d) SiO$_2$. The inset text is the conductivity.  The bottom panels show the corresponding structural models, with Si atoms in brown and O atoms in red. Atomic radii are scaled according to the IPR magnitude, normalized independently for each model, and the gray isosurface represents the KuboMap projection.}
    \label{fig:Figure3}
\end{figure*}

We next examine amorphous silicon-oxide as another test-case for KuboMap. Figure~\ref{fig:Figure3} compares sub-stoichiometric SiO$_x$ models with $x=1.3$, 1.5, and 1.7 [Fig.~\ref{fig:Figure3}(a)--(c)] to stoichiometric SiO$_2$ [Fig.~\ref{fig:Figure3}(d)].  The EDoS and IPR show a clear composition-dependent trend across SiO$_x$. SiO$_{1.3}$ has several near-$E_F$ states with large IPR and the highest KuboMap conductivity (282 S/m). As the oxygen content increases to SiO$_{1.5}$ and SiO$_{1.7}$, the near-\(E_F\) spectral weight decreases and the conductivity drops to $\approx$ 0.3 S/m and 0.02 S/m, respectively. In stoichiometric SiO$_2$, which is a good insulator, the spectrum is depleted around \(E_F\) and the conductivity is strongly suppressed to \(4.6\times10^{-8}\) S/m, consistent with experimentally reported values at $25^\circ$C ($\approx$ 300 K) \cite{Srivastava1985_SiO2}. 

The real-space KuboMap projections show that in sub-stoichiometric SiO$_x$, the transport-active regions are concentrated in Si-rich parts of the network containing connected Si--Si motifs and atoms with large IPR weight. As the oxygen content increases from SiO$_{1.3}$ to SiO$_{1.7}$, these KuboMap regions become progressively less continuous and more spatially confined, consistent with the reduction in conductivity. Stoichiometric SiO$_2$ is qualitatively different: its KuboMap does not form a connected pathway, indicating that although localized states remain, they are not sufficiently connected near $E_F$ to support appreciable transport (similar to $a$-Si-216). Thus, the conductivity decrease from SiO$_{1.3}$ to SiO$_2$ reflects the progressive disruption of the Si-rich conduction network by oxygen incorporation. KuboMap therefore captures the transition from a conducting Si-rich network to an insulating oxide with increasing oxygen content. Additional details for SiO$_x$ are given in Sec. \textcolor{magenta}{S8} \cite{SM}.

In summary, we have introduced KuboMap, a nonnegative real-space conductivity density derived from transport-weighted overlaps of near-Fermi-level Kohn--Sham state pairs. By construction, its spatial integral recovers the total conductivity. KuboMap reveals extended metallic conduction in FCC aluminum, separates insulating and defective amorphous silicon through the presence of connected hopping-like pathways, and tracks the loss of conduction in silicon-oxides as oxygen disrupts Si-rich transport networks. 

Looking ahead, KuboMap’s favorable system-size scaling opens a path toward atomic-to-continuum electric current-flow simulations. This direction parallels recent work by the authors on heat-flow simulations, where the site-projected thermal conductivity (SPTC) method \cite{Ugwumadu2025-SPTC} was used to define the conductivity field for finite element thermal modeling \cite{Ugwumadu2026scacs}. In this context, KuboMap could provide the electronic-transport analogue, enabling spatially resolved electronic conductivity from atomistic simulations to be embedded directly into continuum-scale current-flow simulations.

\begin{acknowledgments}
This work was funded by the Laboratory Directed Research and Development program at Los Alamos National Laboratory (LANL) through a Director’s Postdoctoral Fellowship (20240877PRD4). LANL is operated by Triad National Security, LLC, for the U.S. DOE National Nuclear Security Administration (89233218CNA000001). Data supporting this work are provided at Ref. \cite{Zenodo}.
\end{acknowledgments}

\bibliography{bibliography}

\end{document}


\title{\Large {\bf Supplemental Material for}\\
`Real-Space Mapping of Electronic Conductivity in Complex Materials'}

\author[1]{C. Ugwumadu\thanks{cugwumadu@lanl.gov}}
\author[2]{D. A. Drabold\thanks{drabold@ohio.edu}}
\author[1]{R. M. Tutchton\thanks{rtutchton@lanl.gov}}

\affil[1]{Quantum \& Condensed Matter (T-4) Group, Los Alamos National Laboratory, Los Alamos, NM, USA}
\affil[2]{Department of Physics and Astronomy, Ohio University, Athens, OH, USA}

\date{}

\maketitle

\setstretch{1.5}
\setcounter{suppfigure}{1}

\begin{abstract}
We introduce KuboMap, a real-space representation of electronic conductivity derived from the Kubo--Greenwood formula. KuboMap defines a nonnegative conductivity density whose spatial integral recovers the total conductivity and whose form is guided by Mott's picture of transport through spatially overlapping electronic states. This construction provides a direct map of the transport-active regions of a material. Applied to aluminum, KuboMap recovers an extended metallic conduction network. Applied to amorphous silicon, it distinguishes an insulating defect-free network from a defective structure in which localized near-Fermi states form connected hopping-like pathways. In silicon-oxides, it captures the loss of conduction as increasing oxygen content disrupts Silicon-rich transport networks. KuboMap provides a physically transparent route from Kubo--Greenwood conductivity to real-space transport pathways in complex materials.
\end{abstract}


\section*{Data Availability}
Data supporting this work are provided at Ref. \cite{Zenodo}.

\section{Formal representation of the KuboMap density ansatz}
\label{sec:SM_formal_representation}
We show that the nonnegative density, $\rho_{\alpha \alpha}^{(\eta)}(\mathbf r)$ in Eq.~(\textcolor{magenta}{5}), admits a natural operator representation on a doubled Hilbert space and satisfies the required normalization [Eq. (\textcolor{magenta}{7})] and sum rule [Eq. (\textcolor{magenta}{9})]. This is a formal realization of the ansatz, not a uniqueness proof. The point is that $|\psi_m(\mathbf r)|^2|\psi_n(\mathbf r)|^2$ is not naturally a diagonal kernel on $\mathcal H$, but is one on the doubled space $\mathcal H\otimes\mathcal H$, which is also a Hilbert space, with inner product
\begin{equation}
\langle \phi_1\otimes\psi_1 \mid \phi_2\otimes\psi_2\rangle
= \langle \phi_1\mid \phi_2\rangle\,\langle \psi_1\mid \psi_2\rangle ,
\end{equation}
extended by linearity and completion. Such doubled-space constructions are standard when representing quantities that depend on pairs of states, and closely related ideas appear, for example, in thermo-field dynamics.\cite{Umezawa1985, Roditi1986_TFD}

In the position basis,
\(
\langle \mathbf r|\hat{\Pi}_m|\mathbf r\rangle = |\psi_m(\mathbf r)|^2,
\) where $\hat{\Pi}_m = |m\rangle\langle m|$ projects onto eigenstate $m$, products of diagonal kernels such as $|\psi_m(\mathbf r)|^2|\psi_n(\mathbf r)|^2$ can be written as a single diagonal kernel on $\mathcal H\otimes\mathcal H$\footnote{Following Fano~\cite{Fano1957_DensityMatrix}, one may view operators (e.g.\ the density matrix) as vectors in an $N^2$-dimensional Hilbert--Schmidt space with inner product $\mathrm{Tr}(A^\dagger B)$, which is naturally isomorphic to a doubled space $\mathcal H\otimes\mathcal H^*$ (or equivalently $\mathcal H\otimes\mathcal H$ up to conjugation), thereby justifying the use of tensor-product representations for pair-index quantities such as $|\psi_m(\mathbf r)|^2|\psi_n(\mathbf r)|^2$.} as
\begin{equation}\label{eqn:diagonalKernel}
|\psi_m(\mathbf r)|^2|\psi_n(\mathbf r)|^2
= \langle \mathbf r,\mathbf r \mid (\hat{\Pi}_m\otimes \hat{\Pi}_n)\mid \mathbf r,\mathbf r\rangle ,
\end{equation}
where $|\mathbf r,\mathbf r\rangle := |\mathbf r\rangle\otimes|\mathbf r\rangle$.

We introduce a diagonal resolution on $\mathcal H\otimes\mathcal H$
\begin{equation}\label{eq:Id_diag}
\hat{\hat{\mathbb{I}}}_{\mathrm{diag}} := \int d^3x\; |\mathbf r,\mathbf r\rangle\langle \mathbf r,\mathbf r| ,
\end{equation}
which acts to extract diagonal position kernels\footnote{$\hat{\hat{\mathbb{I}}}_{\mathrm{diag}}$ is not the identity on $\mathcal H\otimes\mathcal H$; traces involving $\hat{\mathbb{I}}_{\mathrm{diag}}$ are understood distributionally and reduce to explicit integrals over $\mathbf r$.}, and from which the overlap matrix can be written compactly as
\begin{align}
\beta_{mn}^{-1}
&= \int d^3x\;\langle \mathbf r|\hat{\Pi}_m|\mathbf r\rangle\,\langle \mathbf r|\hat{\Pi}_n|\mathbf r\rangle \\
&= \mathrm{Tr}\!\left(\hat{\mathbb{I}}_{\mathrm{diag}}\,(\hat{\Pi}_m\otimes \hat{\Pi}_n)\right).
\end{align}
Here $\hat{\mathbb{I}}_{\mathrm{diag}}$ captures the overlap of the diagonal kernels, i.e.\ the overlap of the corresponding probability densities.

Next, for each $(m,n)$ pair, we define the positive operator
\begin{equation}
\widehat{\Xi}_{mn} := \beta_{mn}\,(\hat{\Pi}_m\otimes \hat{\Pi}_n) .
\end{equation}
So that
\begin{align}
\langle \mathbf r,\mathbf r|\widehat{\Xi}_{mn}|\mathbf r,\mathbf r\rangle
&= \beta_{mn}\,\langle \mathbf r|\hat{\Pi}_m|\mathbf r\rangle\,\langle \mathbf r|\hat{\Pi}_n|\mathbf r\rangle \\
&= \beta_{mn}\,|\psi_m(\mathbf r)|^2|\psi_n(\mathbf r)|^2 .
\end{align}
Moreover, the normalization [Eq. (\textcolor{magenta}{7})] is encoded as
\begin{align}
\mathrm{Tr}\!\left(\hat{\mathbb{I}}_{\mathrm{diag}}\,\widehat{\Xi}_{mn}\right)
&=\beta_{mn}\,\mathrm{Tr}\!\left(\hat{\mathbb{I}}_{\mathrm{diag}}\,(\hat{\Pi}_m\otimes\hat{\Pi}_n)\right) 
\equiv 1.
\end{align}

We then define the positive operator
\begin{equation}\label{eq:Sigma_operator}
\widehat{\Sigma}_{\alpha \alpha}^{(\eta)}
:= \sum_{m,n}\gamma_{mn}\,w_m\,w_n\,\widehat{\Xi}_{mn} ,
\end{equation}
and the spatial conductivity density is its diagonal kernel,
\begin{align}
\rho_{\alpha \alpha}^{(\eta)}(\mathbf r)
&:= \langle \mathbf r,\mathbf r|\widehat{\Sigma}_{\alpha \alpha}^{(\eta)}|\mathbf r,\mathbf r\rangle \\
&= \sum_{m,n}\gamma_{mn}\beta_{mn}\,|\psi_m(\mathbf r)|^2|\psi_n(\mathbf r)|^2\,w_m\,w_n ,
\end{align}
which coincides with Eq. (\textcolor{magenta}{5}). Finally,
\begin{align}
\int d^3r\;\rho_{\alpha \alpha}^{(\eta)}(\mathbf r)
&= \mathrm{Tr}\!\left(\hat{\mathbb{I}}_{\mathrm{diag}}\,\widehat{\Sigma}_{\alpha \alpha}^{(\eta)}\right) \\
&= \sum_{m,n}\gamma_{mn}\,w_m\,w_n\,
\underbrace{\mathrm{Tr}\!\left(\hat{\mathbb{I}}_{\mathrm{diag}}\,\widehat{\Xi}_{mn}\right)}_{=\,1} \;
= \sigma_{\alpha \alpha}^{(\eta)} .
\end{align}
as established in Eq. (\textcolor{magenta}{9}).

\section{Geometric interpretation of KuboMap}
\label{sec:SM_geometic_interpretation}

Underlying the discussion of Fig. \textcolor{magenta}{1} in the main text, one may give a simple geometric (real space) interpretation of of the KuboMap conductivity density $\rho (\mathbf r)$ using Mott's picture of the conductivity being the squared of near-Fermi level ($E_F$) states contribution, the \textit{so-called} \(N^2(E_F)\) \cite{Nepal2025_N2}. First, suppose we start from the crude idea that the conductivity is controlled by the density of states at the Fermi level,
\begin{equation}
\sigma \propto N(E_F).
\end{equation}
If \(\psi_m(\mathbf r)\) are Kohn--Sham states with eigenvalues \(E_m\) at position $\mathbf r$, then the corresponding local density of states is
\begin{equation}
\mathcal P_m(\mathbf r)
=
\sum_m \delta(E_m-E_F)\,|\psi_m(\mathbf r)|^2
\end{equation}
If the states near \(E_F\) are localized, this quantity appears as a collection of isolated blobs in space. Regions where $\mathcal P_m(\mathbf r)=0$ contain no near-$E_F$ weight and are inactive in this one-state picture.

The more relevant idea is that what matters is not \(N(E_F)\), but \(\sigma \propto  N^2(E_F)\). This suggests introducing the pair-density field
\begin{equation}
\mathcal{P}_{mn}(\mathbf r)
=
\sum_{mn}
\delta(E_m-E_F)\,
\delta(E_n-E_F)\,
|\psi_m(\mathbf r)|^2\,|\psi_n(\mathbf r)|^2 
\end{equation}
For a pair \((m,n)\), define
\begin{equation}
\mathcal{P}_{mn}(\mathbf r) = \mathcal{P}_{m}(\mathbf r)\;\mathcal{P}_{n}(\mathbf r).
\end{equation}
This product is nonzero only at points in space where both densities are nonzero
\begin{equation}
\mathcal{P}_{mn}(\mathbf r)=0
 \iff
\mathcal{P}_m(\mathbf r)=0
\text{ or }
\mathcal{P}_n(\mathbf r)=0.
\end{equation}

Thus, \(\mathcal{P}_{mn}(\mathbf r)\) identifies where the two states coexist in real space and thereby defines their overlap region $\Omega$,
\begin{equation}\label{eq:App_R_mn}
R_{mn}
=
\left\{
\mathbf r\in\Omega:\mathcal{P}_m(\mathbf r)\mathcal{P}_n(\mathbf r)>0
\right\}.
\end{equation}
In this picture, transport is governed not by individual localized orbitals taken separately, but by the real-space overlap of pairs of states near the Fermi level.
Considering all relevant pairs, labeled by \(\alpha\), each pair defines a set \(R_\alpha\), and the candidate conduction space is the union over all such pairs,
\begin{equation}\label{eq:App_R_union}
\mathcal R=\bigcup_{i=1} R_i.
\end{equation}

This gives a simple geometric criterion: if the regions \(R_\alpha\) remain isolated, \(\mathcal R\) consists of disconnected islands and conduction is hindered; if they form a connected or percolating network, conduction becomes possible.

\section{Numerical implementation of KuboMap}\label{sec:SM_implementation}

The KuboMap field is evaluated on the same real-space grid used to represent the Kohn--Sham orbitals. For a supercell of volume \(\Omega\) and grid dimensions \((n_x,n_y,n_z)\), the number of grid points is
\begin{equation}
N_r=n_xn_yn_z,
\end{equation}
with cell volume
\begin{equation}
\Delta V=\frac{\Omega}{N_r}.
\end{equation}
Continuum integrals are approximated by
\begin{equation}
\int d^3r\,f(\mathbf r)\approx\sum_{i=1}^{N_r}f(\mathbf r_i)\Delta V.
\end{equation}

The overlap matrix is then evaluated on the grid as
\begin{equation}
\beta_{mn}^{-1}
\approx
\sum_{i=1}^{N_r} D_m(\mathbf r_i)D_n(\mathbf r_i)\,\Delta V,  \quad \text{and} \quad
\beta_{mn}
=
\frac{1}{\beta_{mn}^{-1}+\varepsilon}
\label{eq:beta_discrete}
\end{equation}
where
\begin{equation}
D_m(\mathbf r_i)=|\psi_m(\mathbf r_i)|^2
\end{equation}
is the grid-resolved density of band $m$, and $\varepsilon$ is a small regularization parameter. $\beta_{mn}$ is the inverse overlap matrix.

The elements of the momentum matrix are obtained from the wavefunction derivatives as
\begin{equation}
P^{\alpha}_{nm}
=
\sum_{i=1}^{N_r}
\psi_n^*(\mathbf r_i)\,
\partial_\alpha \psi_m(\mathbf r_i)\,
\Delta V,
\qquad \alpha\in\{x,y,z\}
\label{eq:N2_aalpha}
\end{equation}
from which the momentum matrix is constructed as
\begin{equation}
\gamma_{mn}
=
\frac{2\pi e^2\hbar}{\Omega}
\sum_{\alpha=x,y,z}
\left|
-\mathrm{i}\frac{\hbar}{m_e}P^\alpha_{nm}
\right|^2.
\label{eq:N2_gamma_discrete}
\end{equation}

To analyze how the conductivity density is distributed across pairs of Kohn--Sham states, we define the effective pair-strength kernel
\begin{equation}
W_{mn}^{\mathrm{eff}}
=
\gamma_{mn}\beta_{mn}.
\label{eq:W_eff}
\end{equation}
so that the weighted band-pair matrix is then 
\begin{equation}
M_{mn}
=
W_{mn}^{\mathrm{eff}}w_mw_n.
\end{equation}

Analogous to the spectral analysis of the $\Gamma$-matrix in the scape-projected conductivity method \cite{Subedi2020}, we characterize the distribution of the band-pair response by analysis the singular-value spectrum of the effective kernel $W^{\mathrm{eff}}$. We compute the singular value decomposition
\begin{equation}
W^{\mathrm{eff}} = U S V^{T},
\label{eq:svd_main}
\end{equation}
where the diagonal matrix $S$ contains the singular values $s_{\alpha}\ge 0$. For comparison across systems, we consider the normalized spectrum $s_{\alpha}/s_1$, which removes the overall scale of $W^{\mathrm{eff}}$ and isolates the shape of the spectral decay. In this representation, a rapidly decaying spectrum indicates that the response is dominated by a small number of contributing pair modes, whereas a slowly decaying spectrum implies that the response is distributed more broadly across the selected band manifold. The corresponding singular vectors provide the dominant contributing modes associated with these pair-strength channels.

With these definitions, the discrete KuboMap field is
\begin{equation}
\tilde \rho(\mathbf r_i)
=
\sum_{m,n}
M_{mn}\,
D_m(\mathbf r_i)\,
D_n(\mathbf r_i)
\label{eq:N2_final_discrete}
\end{equation}
which is the grid analogue of Eq.~(\textcolor{magenta}{5}). Here, $\mathbf r_i$ labels a grid point, so $\tilde \rho(\mathbf r_i)$ is a scalar field value, not a matrix. Finally, the conductivity is recovered as:
\begin{equation}
\sigma
\approx
\sum_{i=1}^{N_r} \tilde \rho(\mathbf r_i)\,\Delta V
\label{eq:sigma_N2_compact}
\end{equation}

\section{Choice of the broadening parameter}\label{sec:SM_eta}

The Gaussian width, $\eta$ account for both structural finite-size and thermal contributions. So in Eq. (\textcolor{magenta}{10}) we defined
\begin{equation}
   \eta^2 \geq \eta_s^2 + \eta_t^2 ,
\end{equation}
where $\eta_s$ and $\eta_t$ denote the structural and thermal components, respectively. We use \(s\in\{-,+\}\) to denote states below and above \(E_F\), respectively, and write the \(N\) eigenvalues closest to \(E_F\) as \(\{\varepsilon_i^{s}\}_{i=1}^{N}\), ordered by increasing distance from \(E_F\). The successive level spacings are defined as
\begin{equation}
\Delta_i^{s}
=
\begin{cases}
E_F-\varepsilon_1^{-}, & s=-,\ i=1,\\
\varepsilon_{i-1}^{-}-\varepsilon_i^{-}, & s=-,\ i=2,\ldots,N,\\
\varepsilon_1^{+}-E_F, & s=+,\ i=1,\\
\varepsilon_i^{+}-\varepsilon_{i-1}^{+}, & s=+,\ i=2,\ldots,N.
\end{cases}
\label{eq:delta_pm}
\end{equation}

The structural contribution is then estimated as the mean level splitting around the Fermi level,
\begin{equation}
\eta_s
=
\frac{1}{2N}
\sum_{s\in\{-,+\}}
\sum_{i=1}^{N}
\Delta_i^{s}.
\label{eq:eta_s}
\end{equation}
This choice is consistent with the finite-size argument of Allen and Feldman~\cite{AllenFeldman1993} for thermal conductivity, where they emphasized that in finite periodic (disordered) structural models, the spectral broadening must exceed the relevant mean level spacing of the vibrational eigenfrequencies.

The thermal contribution is estimated from fluctuations of the same near-\(E_F\) manifolds along the MD trajectory. For each snapshot at time \(t_j\), we define
\begin{equation}
\bar{\varepsilon}^{s}(t_j)
=
\frac{1}{N}
\sum_{i=1}^{N}
\left[
\varepsilon_i^{s}(t_j)-E_F(t_j)
\right],
\quad s\in\{-,+\}.
\label{eq:epsbar_pm}
\end{equation}

The thermal broadening is estimated as the average root-mean-square fluctuation of these two branches \cite{Drabold1991},
\begin{equation}
\eta_t
=
\frac{1}{2}
\sum_{s\in\{-,+\}}
\sqrt{
\frac{1}{M}
\sum_{j=1}^{M}
\left(
\bar{\varepsilon}^{s}(t_j)
-
\left\langle \bar{\varepsilon}^{s} \right\rangle_t
\right)^2},
\label{eq:eta_t}
\end{equation}
where \(M\) is the number of equilibrated MD snapshots and \(\langle\cdot\rangle_t\) denotes the corresponding time average.

\begin{figure}[!t]
    \centering
    \includegraphics[width=\linewidth]{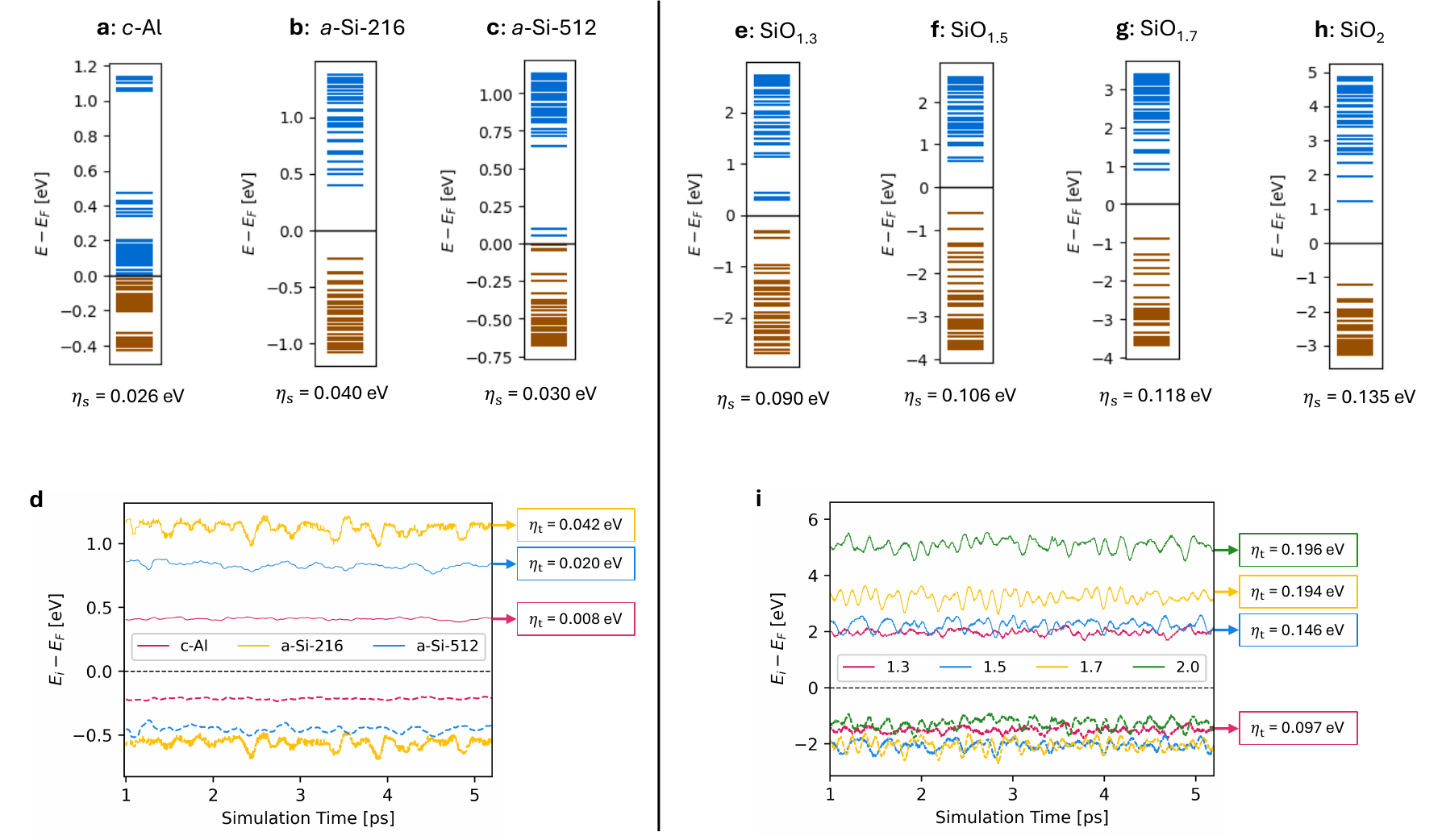}
    \caption{Determination of the Gaussian broadening threshold $\eta$. The structural contribution, $\eta_s$, is estimated from the 30 Kohn--Sham states above (blue) and below (brown) the Fermi level (black dashed line) for (a) $c$-Al, (b) $a$-Si-216, and (c) $a$-Si-512. The corresponding thermal contribution, $\eta_t$, obtained from MD at 300 K, is shown in (d). Panels (e)--(h) show the $\eta_s$ analysis for the silicon-oxides SiO$_{1.3}$, SiO$_{1.5}$, SiO$_{1.7}$, and SiO$_{2}$, respectively, while panel (i) shows their corresponding $\eta_t$ values from MD at 300 K.}
    \label{fig:Sfig_eta_eval}
    \refstepcounter{suppfigure}
\end{figure}

Figure~\ref{fig:Sfig_eta_eval} show the evaluation of the structural and thermal contributions to the Gaussian broadening parameter \(\eta\) for all models. Figure~\ref{fig:Sfig_eta_eval}(a)--(c) show the near-\(E_F\) Kohn--Sham levels used to estimate the structural contribution, \(\eta_s\), from 30 states below and 30 above the Fermi level for thermally disordered \(c\)-Al, \(a\)-Si-216, and \(a\)-Si-512, respectively, while Fig.~\ref{fig:Sfig_eta_eval}(d) shows the corresponding root-mean-square fluctuations used to determine the thermal contribution, \(\eta_t\). Figure~\ref{fig:Sfig_eta_eval}(e)--(h) present the corresponding \(\eta_s\) analysis for the SiO\(_x\) series, SiO\(_{1.3}\), SiO\(_{1.5}\), SiO\(_{1.7}\), and SiO\(_2\), and Fig.~\ref{fig:Sfig_eta_eval}(i) shows the corresponding \(\eta_t\) values obtained from fluctuations of the same near-\(E_F\) states along the 300 K MD trajectory.

\section{Room-temperature conductivity estimate for the structures}\label{sec:SM_conductivity_EF}
Table~\ref{tab:conductivity_EF} summarizes the conductivity for all snapshots obtained at 300 K for all the models considered. The structure files (in VASP \footnote{Vienna \textit{ab initio} Simulation Package} format) used in this work are provided as supplementary data \cite{Zenodo}. 

For crystalline Al, the conductivity is relatively stable across the ten snapshots, with an average value of \(3.5\times10^{5}\) S/m. The small snapshot-to-snapshot variation reflects the character of the thermal broadening $\eta_t$ in Fig. \ref{fig:Sfig_eta_eval}(d). 

\begin{table*}[!h]
\refstepcounter{supptable}
\renewcommand{\arraystretch}{1.2}
\caption{Conductivity \(\sigma\) (in S/m) for all snapshots at 300 K, for all the systems.}
\small
\label{tab:conductivity_EF}
\centering
\begin{threeparttable}
\begin{tabular*}{\linewidth}{@{\extracolsep{\fill}}c|ccccccc}
\hline
Snapshot
& $c$-Al\tnote{a}
& $a$-Si-216\tnote{b}
& $a$-Si-512
& SiO$_{1.3}$
& SiO$_{1.5}$\tnote{c}
& SiO$_{1.7}$\tnote{d}
& SiO$_2$\tnote{e} \\
\hline
S1  & 3.15 & 275  & 806 & 182 & 6.75 & 1.66 & 257 \\
S2  & 2.65 & 1463   & 527 & 195 & 6.05 & 3.79 & 135 \\
S3  & 3.46 & 5.40 & 320 & 161 & 7.47 & 1.02 & 7.55 \\
S4  & 2.80 & 5.46 & 250 & 851 & 3.40 & 1.93 & 1.79 \\
S5  & 3.65 & 800  & 435 & 196 & 40.3 & 1.97 & 12.0 \\
S6  & 5.08 & 5.72 & 506 & 162 & 42.4 & 1.55 & 16.5 \\
S7  & 4.24 & 5.90 & 394 & 140 & 8.25 & 1.44 & 0.13 \\
S8  & 2.67 & 6.91 & 630 & 147 & 8.99 & 2.64 & 4.85 \\
S9  & 4.21 & 6.36   & 743 & 237 & 8.49 & 1.96 & 0.42 \\
S10 & 3.03 & 6.62 & 369 & 143 & 130  & 1.56 & 24.9 \\
\hline
Avg. & 3.49 & 258 & 498 & 241 & 26.2 & 1.95 & 46.0 \\
\hline
\end{tabular*}
\begin{tablenotes}
\footnotesize
\item[a] For \(c\)-Al, \(\sigma\) is reported in units of \(10^{5}\) S/m.
\item[b] For \(a\)-Si-216, \(\sigma\) is reported in units of \(10^{-7}\) S/m.
\item[c,d] For SiO$_{1.5}$ and SiO$_{1.7}$, \(\sigma\) is reported in units of \(10^{-2}\) S/m.
\item[e] For SiO$_2$, \(\sigma\) is reported in units of \(10^{-7}\) S/m.
\end{tablenotes}
\end{threeparttable}
\end{table*}

The amorphous Si models show stronger sensitivity to atomic configuration. For $a$-Si-216, most snapshots have very small conductivities, while a few exhibit much larger values, giving an average conductivity of \(\approx 2.6\times10^{-5}\) S/m. In large-gap amorphous systems such as (a)-Si-216, transport is expected to occur only when thermal motion brings localized states into favorable energetic and spatial alignment \cite{Anderson1958}.

Previous time-dependent Kohn--Sham simulations suggest that hopping in amorphous Si requires timescales of order 100 fs \cite{Jun2003}. We therefore exclude transient spectral coincidences that are too short-lived to contribute meaningfully to the conductivity estimate. This timescale criterion was considered when selecting representative \(a\)-Si-216 snapshots for conductivity evaluation. To illustrate this behavior, we provide an animation of a 5 ps \(a\)-Si-216 trajectory as supplementary data \cite{Zenodo}, showing Fermi-gap fluctuations and the lifetimes of configurations that may support hopping transitions. In contrast, \(a\)-Si-512 has a much larger average conductivity of 498 S/m, consistent with coordination defects and more persistent near-\(E_F\) states that form connected transport pathways.

The SiO$_x$ models exhibit a systematic reduction in conductivity with increasing oxygen content. SiO$_{1.3}$ has the largest average conductivity among the oxide compositions, with an average value of 241 S/m, reflecting the presence of a more connected Si-rich network. As the oxygen concentration increases to SiO$_{1.5}$ and SiO$_{1.7}$, the average conductivity decreases to \(26.2\times10^{-2}\) and \(1.95\times10^{-2}\) S/m, respectively. For stoichiometric SiO$_2$, the conductivity is strongly suppressed, with an average value of \(46.0\times10^{-7}\) S/m, consistent with its insulating character \cite{Srivastava1985_SiO2}.

\section{Additional analysis of crystalline aluminum}\label{sec:SM_cAl}

\begin{figure}[!h]
    \centering
    \includegraphics[width=\linewidth]{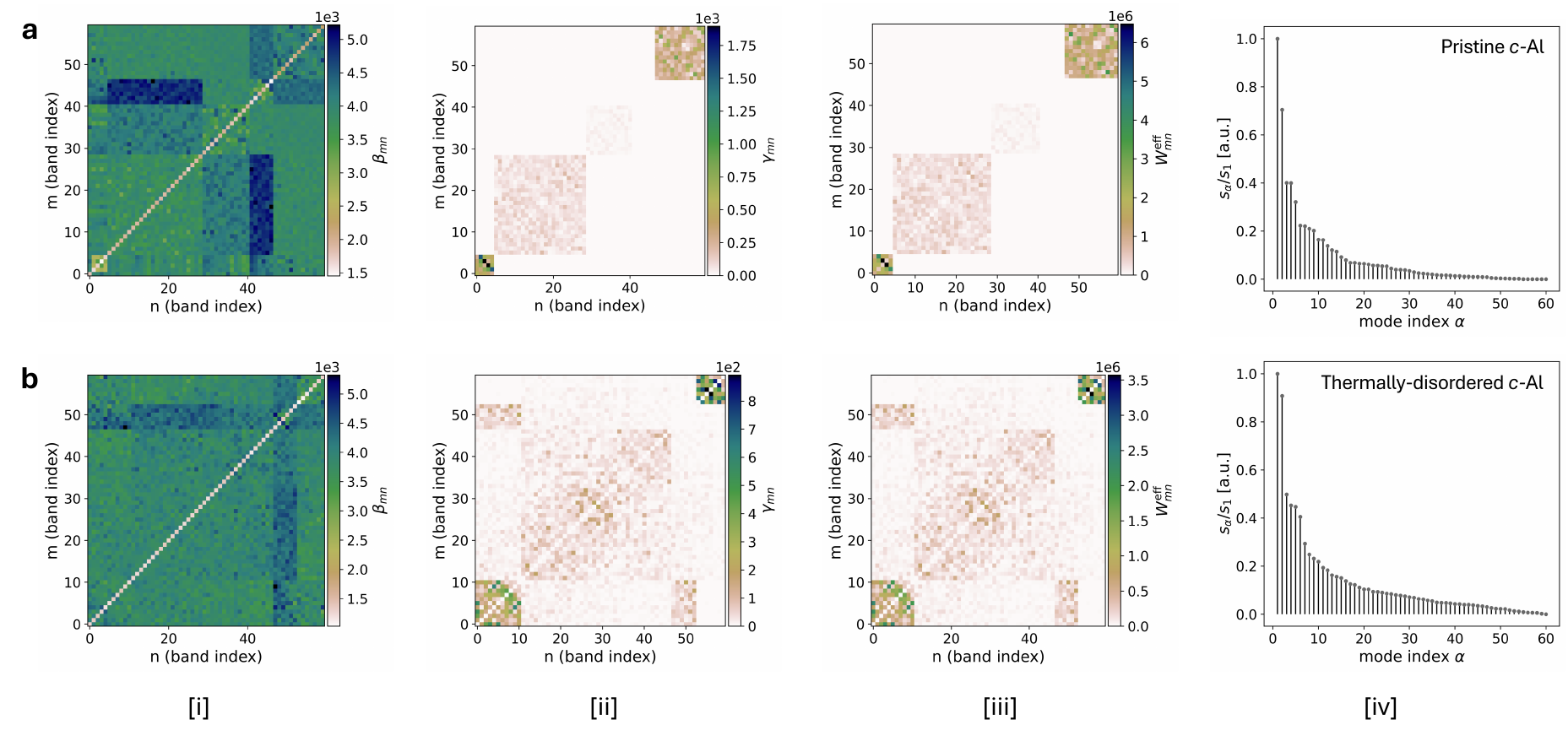}
    \caption{KuboMap Analysis for (a) pristine $c$-Al and (b) thermally-disordered $c$-Al. Colormaps of [i] the inverse overlap matrix, $\beta_{mn}$; [ii] the momentum matrix, $\gamma_{mn}$; and [iii] the corresponding unweighted effective pair-strength kernel, $W_{mn}^{\mathrm{eff}}$, shown for 30 states below and 30 states above the Fermi level, located at band index 30. [iv] Normalized singular-value spectrum of $W^{\mathrm{eff}}$.}
    \label{fig:Sfig_Al}
    \refstepcounter{suppfigure}
\end{figure}

Figure~\ref{fig:Sfig_Al} compares the band-pair structure of the (a) pristine $c$-Al and the thermally disordered $c$-Al. Figures~\ref{fig:Sfig_Al}(a,i) and \ref{fig:Sfig_Al}(b,i) shows the inverse overlap matrix, $\beta_{mn}$ [Eq. \eqref{eq:beta_discrete}], evaluated for 30 states below and 30 states above the Fermi level. In this indexing convention, \(E_F\) is located at band index 30, with indices numbered from 0 to 60. The $\beta_{mn}$ matrix exhibits a strong diagonal structure, indicating that each state has the largest self-overlap, as expected. However, the off-diagonal entries are also appreciable and broadly distributed, reflecting the extended metallic character of the near-$E_F$ states in Al. 

The momentum matrix, $\gamma_{mn}$, shown in Fig.~\ref{fig:Sfig_Al}(a,ii) for the pristine $c$-Al is considerably more sparse than $\beta_{mn}$. This indicates that spatial overlap alone does not determine the transport response; only state pairs with appreciable velocity, or momentum, matrix elements contribute significantly to the conductivity. The largest $\gamma_{mn}$ values are confined to specific band-pair regions, showing that transport in pristine crystalline Al is dominated by selected couplings within the near-Fermi manifold. By contrast, in thermally disordered $c$-Al [Fig.~\ref{fig:Sfig_Al}(b,ii), nonzero $\gamma_{mn}$ values are more broadly distributed across the off-diagonal elements, indicating that thermal disorder relaxes the symmetry-imposed selection of band-pair couplings and spreads the momentum response over a larger set of near-Fermi states.

The effective pair-strength kernel, $W_{mn}^{\mathrm{eff}}$ [Eq. \eqref{eq:W_eff}], shown in the third column for both systems [Fig.~\ref{fig:Sfig_Al}(a/b;iii)], combines the spatial-overlap and momentum-coupling information. The resulting kernel retains the sparsity of $\gamma_{mn}$ while weighting the transport-active pairs by their real-space overlap. Thus, $W_{mn}^{\mathrm{eff}}$ identifies the subset of near-$E_F$ state pairs that dominate the KuboMap response. 

The low-rank structure of the effective pair-strength kernel is further demonstrated by the normalized singular-value spectrum in Fig.~\ref{fig:Sfig_Al}(a/b;iv). The spectrum decays rapidly with mode index, indicating that the dominant transport response can be represented by a relatively small number of modes. Physically, this suggests that although many state pairs are included in the near-$E_F$ window, the effective conductivity density is controlled by a smaller subset of collective pairwise contributions. The decay is more rapid in pristine $c$-Al where states above mode index ``40" tend to zero.

\section{Additional analysis of amorphous silicon}
\label{sec:SM_aSi}

\begin{figure}[!h]
    \centering
    \includegraphics[width=\linewidth]{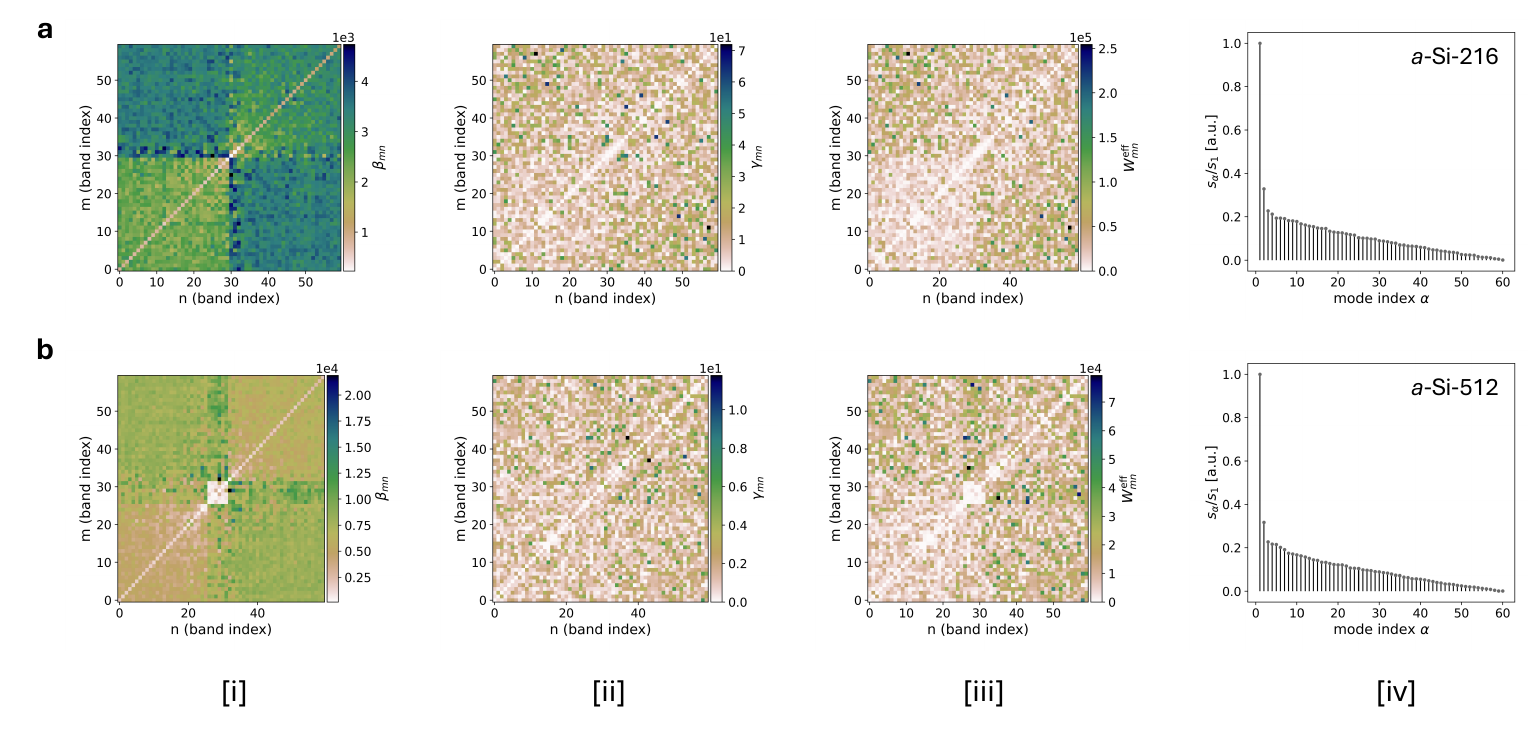}
    \caption{KuboMap Analysis for (a) $a$-Si-216 and (b) $a$-Si-512. Colormaps of [i] the inverse overlap matrix, $\beta_{mn}$; [ii] the momentum matrix, $\gamma_{mn}$; and [iii] the corresponding unweighted effective pair-strength kernel, $W_{mn}^{\mathrm{eff}}$, shown for 30 states below and 30 states above the Fermi level, located at band index 30. [iv] Normalized singular-value spectrum of $W^{\mathrm{eff}}$.}
    \label{fig:Sfig_Si}
    \refstepcounter{suppfigure}
\end{figure}

Figure~\ref{fig:Sfig_Si} compares the band-pair structure of the two amorphous Si models, $a$-Si-216 and $a$-Si-512. For $a$-Si-216, $\beta_{mn}$ shows broad but structured band-pair correlations, with a visible change across the Fermi-level band index ("30"). This reflects the separation between valence- and conduction-side states in the nearly defect-free network. In contrast, the momentum matrix $\gamma_{mn}$ is more uniformly distributed and lacks strongly localized high-intensity blocks, indicating that no small subset of state pairs produces a dominant transport channel. The corresponding $W_{mn}^{\mathrm{eff}}$ retains this broadly distributed character, consistent with the absence of a connected real-space KuboMap pathway in the main text. The singular-value spectrum decreases gradually after the leading mode, suggesting that the weak residual response is spread over many small pair contributions rather than concentrated in a few dominant transport channels.

The $a$-Si-512 model shows a qualitatively different overlap structure. Here, $\beta_{mn}$ contains stronger localized features near the Fermi-level band index, consistent with the presence of defect- or tail-derived states close to $E_F$. These near-Fermi states have enhanced spatial localization and therefore produce sharper pair-overlap signatures. The momentum matrix $\gamma_{mn}$ remains relatively sparse and irregular, but the corresponding $W_{mn}^{\mathrm{eff}}$ shows that a subset of near-Fermi state pairs has sufficient overlap and momentum coupling to contribute to transport. This supports the  interpretation that conduction in $a$-Si-512 is defect-mediated and associated with localized states that form a connected hopping-like pathway. The singular-value spectra for both amorphous Si models decay more slowly than in crystalline Al, indicating that the transport response is less dominated by a few symmetry-selected metallic modes. Instead, the amorphous systems distribute their band-pair response over a broader set of pair channels.

\section{Additional analysis of silicon-oxides}
\label{sec:SM_SiOx}

\begin{figure}[!h]
    \centering
    \includegraphics[width=\linewidth]{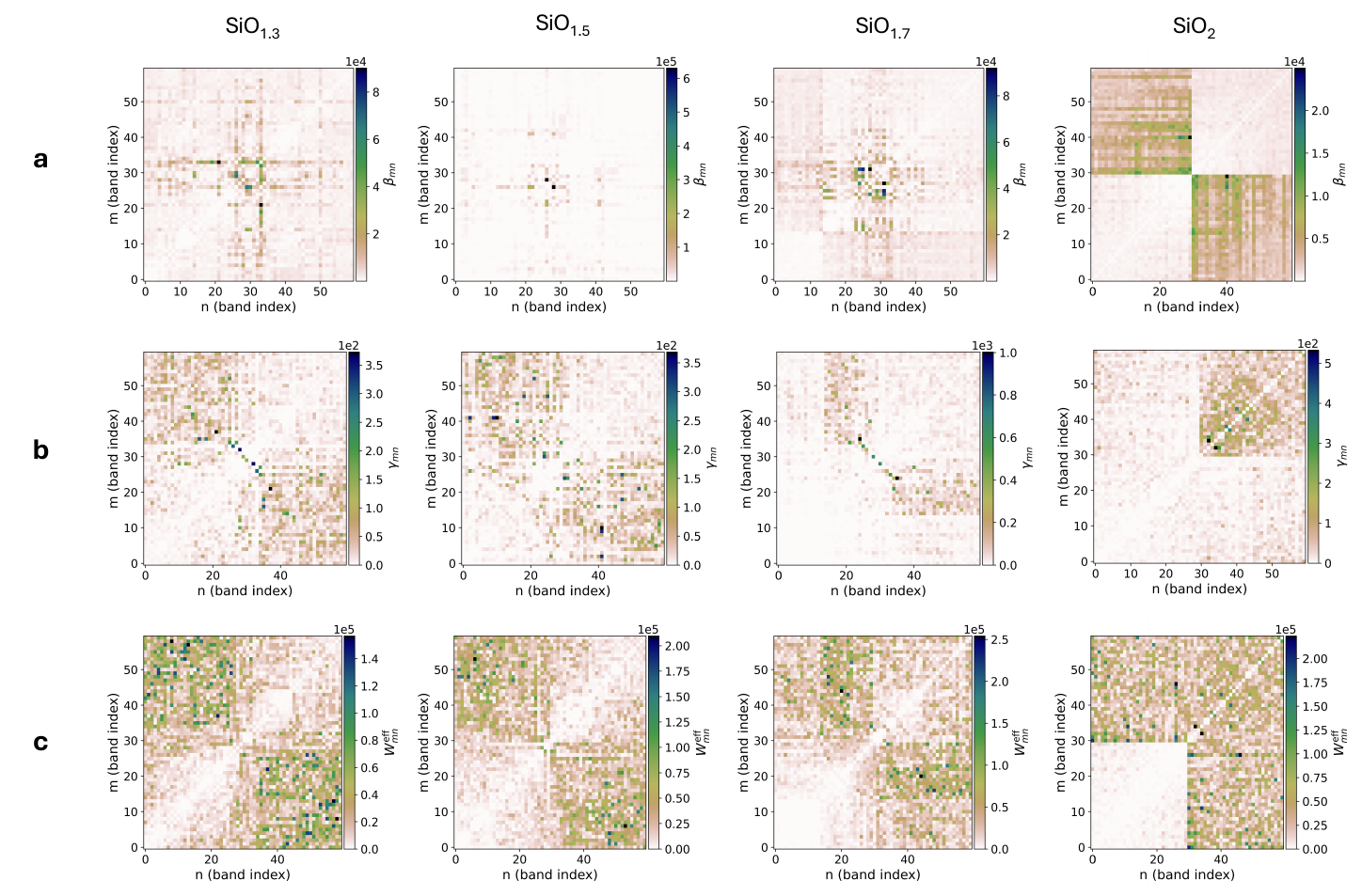}
    \caption{Analysis of Si--O systems. Colormaps of (a) the inverse overlap matrix $\beta_{mn}$, (b) the momentum matrix, $\gamma_{mn}$; and (c) the corresponding unweighted effective pair-strength kernel, $W_{mn}^{\mathrm{eff}}$, shown for 30 states below and 30 states above the Fermi level, located at band index 30. for SiO$_\mathrm{1.3}$, SiO$_\mathrm{1.5}$, SiO$_\mathrm{1.7}$, and SiO$_\mathrm{2}$ (column-wise).}
    \label{fig:Sfig_SiO}
    \refstepcounter{suppfigure}
\end{figure}

Figure~\ref{fig:Sfig_SiO} shows the inverse overlap matrix $\beta_{mn}$, momentum matrix $\gamma_{mn}$, and effective pair-strength kernel $W_{mn}^{\mathrm{eff}}$ for the silicon-oxides. The sub-stoichiometric models, SiO$_{1.3}$, SiO$_{1.5}$, and SiO$_{1.7}$, exhibit unique, unstructured band-pair patterns, compared to stoichiometric SiO$_{2}$. 

The inverse overlap matrices in Fig.~\ref{fig:Sfig_SiO}(a) show that the sub-stoichiometric systems retain appreciable pair overlap among states near the Fermi level. For SiO$_{1.3}$ and SiO$_{1.7}$, the overlap is concentrated around the near-$E_F$ band indices, with cross-like features extending across the valence- and conduction-side states. SiO$_{1.5}$ shows a more localized overlap pattern, suggesting that fewer state pairs dominate the near-Fermi overlap. In contrast, SiO$_2$ exhibits a more block-separated overlap structure, reflecting the stronger separation between occupied and unoccupied manifolds in the insulating oxide.

The momentum matrices in Fig.~\ref{fig:Sfig_SiO}(b) are generally more selective than the inverse overlap matrices. In the sub-stoichiometric systems, nonzero $\gamma_{mn}$ values appear primarily in band-pair regions associated with near-Fermi states, indicating that the Si-rich networks provide both spatial overlap and current-carrying coupling, as they form the states near $E_F$ (see main text). The SiO$_{1.7}$ case is particularly sparse, suggesting that although overlap exists, the number of strongly current-active pairs is reduced. For SiO$_2$, the momentum response is concentrated mainly in a separated block of states, consistent with the absence of a connected near-Fermi conduction network.

The effective kernels in Fig.~\ref{fig:Sfig_SiO}(c) combine these two requirements and identify the band pairs that are most relevant to the KuboMap conductivity. The sub-stoichiometric models retain significant $W_{mn}^{\mathrm{eff}}$ structure, especially in regions coupling states across the selected near-Fermi window. These patterns are consistent with the finite conductivities reported in the main text and with the real-space KuboMap projections showing conduction-active Si-rich regions. As oxygen content increases, the effective pair response becomes more fragmented, reflecting the progressive disruption of the Si-derived transport network by oxygen incorporation. Stoichiometric SiO$_2$ shows a qualitatively different kernel, dominated by separated blocks rather than a connected near-Fermi band-pair response, consistent with its strongly suppressed conductivity.

\section*{Supplemental references}
\bibliography{bibliography}